\documentclass[10pt,nocopyrightspace]{sigplan-proc-varsize-package/sigplan-proc-varsize}

\usepackage{graphicx,latexsym,cite,setspace,algorithmic,subfigure,epsfig,float,comment,url,setspace} 

\numberofauthors{2}

\author{
\alignauthor Vinayak Naik \\
        \affaddr{BITS Pilani, Goa}\\
       \email{vinayak@goa.bits-pilani.ac.in}
\alignauthor Anish Arora\\
        \affaddr{The Ohio State University}\\
       \email{anish@cse.ohio-state.edu}
 }
 
\begin{spacing}{1.15} 
 
\title{Harvest: A Reliable and Energy Efficient Bulk Data Collection Service for Large Scale Wireless Sensor Networks} 

\begin{document}

\maketitle

\newtheorem{theorem}{Theorem}
\newtheorem{conjecture}[theorem]{Conjecture}

\large

\begin{abstract}
\large{\emph{We present a bulk data collection service, Harvest, for energy constrained wireless sensor nodes. To increase spatial reuse and thereby decrease latency, Harvest performs concurrent, pipelined exfiltration from multiple nodes to a base station. To this end, it uses a distance-k coloring of the nodes, notably with a constant number of colors, which yields a TDMA schedule whereby nodes can communicate concurrently with low packet losses due to collision. This coloring is based on a randomized CSMA approach which does not exploit location knowledge. Given a bounded degree of the network, each node waits only O$(1)$ time to obtain a unique color among its distance-k neighbors, in contrast to the traditional deterministic distributed distance-k vertex coloring wherein each node waits O$(\Delta^{2})$ time to obtain a color.}}

\emph{Harvest offers the option of limiting memory use to only a small constant number of bytes or of improving latency with increased memory use; it can be used with or without additional mechanisms for reliability of message forwarding.}

\emph{We experimentally evaluate the performance of Harvest using 51 motes in the Kansei testbed. We also provide theoretical as well as TOSSIM-based comparison of Harvest with Straw, an extant data collection service implemented for TinyOS platforms that use one-node at a time exfiltration. For networks with more than 3-hops, Harvest reduces the latency by at least 33\% as compared to that of Straw.} 
\end{abstract}

\vspace{.15cm}
\section{Introduction}

Wireless sensor nodes often maintain logs of network, environment, middleware, and application behavior.  Examples of logged information include link qualities, network routes, sensory data, mobility traces, exception reports, application statistics, etc.  The collection of bulk data from a number of wireless sensor nodes is thus a frequent requirement for testers, operators, managers, modelers, and users.   In this paper, we focus on the convergecast of the (potentially different) bulk data logged at a number of nodes to one ``base station'' node. We consider an ``off-line'' setting where no other data traffic is present on the network; this case arises when the bulk data collection can materially interfere with ongoing application traffic or when the size/generation-rate of the bulk data exceeds the effective communication capacity of the source nodes with respect to the base station.

As networks scale to larger numbers of nodes and communication hops and as the bulk datum sizes grow, the reliability, energy-efficiency, and latency of the collection operation become key issues. While these issues have been well studied in the context of bulk data dissemination, they have received far less attention in the case of bulk data convergecast. Also, since network debugging and management are primary motivations for the collection service, it is desirable that this service have a small footprint in terms of instruction memory, data memory, message overhead, and wireless traffic, and to minimize its dependence on other network services, including localization and time synchronization. In this paper, we present and evaluate a protocol that meets these requirements; we call this protocol Harvest.

Harvest achieves its reliability with two measures: First, it schedules the transmissions (of messages containing bulk datum pieces) so that message losses due to collision are reduced. We reduce the problem of computing a TDMA schedule that avoids hidden terminals to computing a distance-2 (henceforth D-2) vertex coloring; the color of a node decides the slot in which it can transmit. In the unit disk graph model, these two problems are equivalent \cite{sk:channel_models}: this intuitively follows from the observation if two non-neighboring nodes $u$ and $v$ interfere with each other then there exists a node $w$ such that there are edges $(u,w)$ and $(w,v)$ in the unit disk graph. Second, Harvest uses acknowledgments and retransmissions at the MAC layer for recover from losses 

Harvest achieves its energy efficiency by avoiding energy-intensive flash operations: it performs at most one flash read and no flash write for any bulk datum piece, and stores the piece at nodes en route to the base station only in their RAM. Of course, avoiding message collision also yields energy efficiency. Harvest keeps the message overhead low (below 9 bytes per bulk data packet). To keep the number of control message tranmissions low, we present a distributed algorithm for its TDMA schedule computation that generates O(1) control message per node involved, which is significantly better than the traditional, distributed alternative, which incurs O$(\Delta^{2})$ per node involved, where $\Delta$ is the node degree. As we explain shortly, this improvement is enabled by computing vertex coloring with a constant number of colors that may be smaller than $\Delta$. As a result, not every node in the network is colored and so the TDMA schedule has to be computed in an ongoing manner, which in turn implies the control message savings is an ongoing one. Moreover, since an idle radio also consumes significant power (of the same order as that during message reception), Harvest schedules the switching off of the radio to save energy: asymptotically, a node keeps its radio on only for the time it is scheduled to send data on behalf of itself or someother node.

Finally, Harvest achieves low latency in two ways. For one, it exploits spatial reuse. Instead of collecting data from only one node at a time, Harvest allows data collection from some constant number of nodes concurrently. For ease of exposition only, we let the concurrency constant be 2 henceforth (the protocol assumes that the user will specify this concurrency constant as a parameter; in fact, larger constants for dense networks yield lower latency). For nodes (including the base station) to concurrently receive data from 2 nodes, 4 colors are needed in the vertex coloring (one for the node, two for its allowed children, and one for its parent node). In other words, regardless of the density of the network, Harvest colors at most 4 nodes in the inteference region of any node at any time. Once a colored node has completed its transmissions, Harvest lets an uncolored node in its interference region to assume that color; this is the on-going aspect of the TDMA schedule computation. We validate that a concurrency constant of 2 yields 33\% less latency for large networks ---specifically networks larger than 3 hops--- than the approach that collects data from one node at a time; the latter approach is adopted by the Straw protocol.  This latency improvement is obtained even if we restrict Harvest to allow concurrent reception at only the base station and reception from at most one child node at every other colored node.
And two, we ensure that the control algorithms used by Harvest have constant time complexities. In particular, our TDMA slot synchronization algorithm (which obviates the need for a time synchronization service) has a local convergence time of O$(1)$ as opposed to O$(\Delta^{2})+$O$(D)$ in traditional algorithms, given the bound on the node density. 

\vspace{0.15cm}
\noindent
\textbf{Contributions of the paper}.~~~
\begin{enumerate}
	\item  We present a randomized distributed algorithm that assigns a constant number of colors in the D-2 region of every node (if such a coloring is possible) so that each node that gets a color waits O$(1)$ time until it gets a unique color; this contrasts with the O$(\Delta^2)$ wait time of traditional deterministic D-2 coloring algorithms. This is achieved by executing our TDMA scheduling algorithm on top of a CSMA/CA-based MAC; thus, in the event that two nodes within reliable range of each other will start to contend on a color simulataneously (i.e. in the same slot), then due to CSMA/CA, one of the two nodes will back off and receive a packet from the other contending node, therby yielding that color. We note that the algorithm would work given (local or nonlocal) ways of calculating the interference regions of nodes other than our method for local computation of the D-2 neighborhood set.

	\item We present an algorithm that synchronizes its TDMA slots with that other colored nodes within O$(1)$ time of its being colored. Traditional deterministic TDMA slot synchronization algorithms have O$(Dia)$ convergence time, where $Dia$ is the diameter of the network. Intuitively, the reason for the O$(Dia)$ convergence time is that the maximum number of colors used the nodes in any interference region needs to be propagated to all the nodes after the D-2 coloring so that they can agree on the transmission period. Harvest achieves constant time converence because of its use of a pre-specified constant number of colors.

	\item We present a data collection algorithm that can use a small constant number (even 1) of packet-sized buffers, irrespective of the number of nodes and the bulk datum sizes. In effect, each node in Harvest can have at the most two children node that send data to it. Using 1 buffer let's the node forward on behalf of only one child; using two let's the node forward on behalf of both; and using more buffers helps in further reducing the latency. 

	\item We evaluate the performance benefit of the spatial reuse and TDMA scheduling in Harvest, by providing a comparative performance with the Straw protocol, in terms of latency and energy improvement achieved via the former and the relative message overheads (Straw use 7 bytes per packet versus Harvest's 9). 
\end{enumerate}
 
\vspace{0.15cm}
\noindent
\textbf{Organization of the paper}.~~~In Section 2, we present the system model and problem statement in more detail. We describe the Harvest protocol and its TinyOS implementation in Section 3. We analyse the performance of the randomized TDMA scheduling algorithm and the data collection components of Harvest in Section 4. In Section 5, we overview the Straw protocol and compare its performance with that of Harvest, analytically and via TOSSIM simulations. We describe a number of extensions of Harvest and discuss its relevance for collecting on-line streaming data in Section 6. We discuss related work in Section 7 and make concluding remarks in Section 8.


\vspace{0.15cm}
\section{The Bulk Data Convergecast Problem}

The system consists of $n, n>0,$ wireless sensor nodes, called motes, one of which is distinguished as a base station. We do not make assumptions of how the mote locations are spatially distributed, nor do we assume the availability of a location service. We do assume that each mote can communicate with the base station over zero or more communication hops and the degree $\Delta$ in the network is bounded.

Some of the motes initially each have a bulk datum in their data store, whose size may vary from mote to mote and may exceed that of the mote's RAM. For simplicity, we assume that each bulk datum resides in the nonvolatile store of its mote. Desired is a middleware service that upon initiation from the base station collects of all motes' bulk data at the base station. In decreasing order of importance, the performance metrics of the service are: first, energy efficiency and reliability, and, second, latency, by which we mean the time taken at the base station from the initiation to the completion of the bulk data collection opreation.  

With respect to energy, Table \ref{table:current_draw} illustrates the energy cost in terms of current draw for common operations for the case of motes in the Mica-2/XSM family.
\begin{table}[htp]
\centering
\begin{tabular}{|c|c|} \hline 
\textbf{Operation}		&	\textbf{Current Draw} 	\\ \hline
CPU and Idle Radio		&	8 mA 				\\ 
Packet Reception			&	7.03 mA 			\\
Packet Transmission		&	10.4 mA 			\\
EEPROM Read				&	6.2 mA 			\\
EEPROM Write			&	18.4 mA 			\\ \hline
\end{tabular}
\vspace{0.15cm}
\caption{Energy required by common operations}
\label{table:current_draw}
\end{table}

The table suggests that given the aggregate current draw for an EEPROM read and write (24.6mA) is significant, and thus every addition of flash operations to the bulk convergecast forwarding process will almost double the current draw associated with the minimum aggregate current draw of radio receive, CPU, and radio transmission (25.43mA). It follows that minimizing EEPROM operations is desirable for the energy metric. The table also identifies the energy overhead associated with an idle radio. One implication is that a mote should sleep as soon as it has no data to send of its own or on behalf of other motes.

With respect to reliability, we focus attention on obtaining high but not necessarily 100\% reliable data collection. Unlike the dual problem of bulk data dissemination, where objectives such as mote reprogramming demand all-or-nothing delivery of bulk data, the use cases of bulk data convergecast can often tolerate low levels of unreliability. In designing our solution, we do not emphasize a particular selection of a link estimation technique or retrasmission mechanism. (Specifically, our experimental evaluation of our solution uses the WMEWMA link estimation approach of Woo and Culler \cite{aw:mintroute} and 0-retransmissions, but these choices are not of central importance.)

With respect to latency, we note that the problem statement does not emphasize the latency of collection from the perspective of individual motes. Had we considered the version of the problem where motes were continually generate and stream data to the base station, low jitter and comparable latency across the motes would have been desirable. We therefore regard these latter requirements as being optional, but not first order, for solving the problem.

Finally, in designing our solution, we do not assume the availability of a time synchronization service.

\vspace{0.15cm}
\section{The Harvest Protocol}

\vspace{0.15cm}
\subsection{The Components of Harvest}

In this section, we describe the three components of Harvest, viz. interference neighborhood discovery, randomized slot assignment and synchronization, and data collection.
\vspace{0.15cm}

\label{'section:harvest_protocol'}

\noindent 
\textbf{Neighborhood Discovery.}~~~Each node performs online link estimation to find out its 1-hop neighborhood set. For ease of exposition, we first assume that all the links are symmetric, i.e., the link quality between two nodes is same in both the directions. Therefore, it is sufficient to do link estimation in any one direction. However, the links in sensor network may not always be symmetric, so we will extend the link estimation in both the directions to deal with asymmetric links. 

A number of metrics can be used for this link estimation; for instance we may use the window mean with exponentially weighted moving average (WMEWMA) metric. This metric was has been used by MintRoute protocol \cite{aw:mintroute} of Woo and Culler. There are two tuning parameters for WMEWMA-based link estimations, viz. $\alpha$ and $t$. The parameter $\alpha$ determines the size of the history used in link estimation and $t$ determines the rate at which link estimation is updated.(Experimental results in the literature show that the values 0.6 and 30 for $\alpha$ and $t$ respectively, provide stable and agile link estimation for Chipcon's CC1000 radio. In particular, the settling time, which is the length of time for the estimator to converge within $\pm10\%$ of the actual value and remain within the error bound.)

Based on link estimation, path selection to the base station can be based again on a number of metrics studied in the literature, e.g., end-to-end path reliability, hop distance, end-to-end mac latency, etc;. for instance, we may use a combination of link quality and hop distance. In particular, we define the 1-hop neighborhood of a node $A$ to be the set of nodes that have WMEWMA value greater than or equal to 75 (which roughly implies a stable packet loss rate less than 10\%). Among the 1-hop neighbors, node $A$ selects a node with the least hop distance to the base station as its parent. Using TOSSIM simulations, we find that the minimum WMEWMA link quality between two nodes at 2-hops from each other is 30 (which roughly implies the nodes can reliably sense each other's carrier). 
\vspace{0.15cm}


\noindent 
\textbf{Randomized Slot Assignment and Synchronization.}~~~As explained in Section 1, Harvest uses 4 colors in the entire network (i.e., two more colors than our de facto concurrency constant of 2). The TDMA scheduling divides time into intervals of length $T = 4*t_{S}$, where $t_{S}$ is the duration of a timeslot. Note that the color assignment should be such two nodes with that are not within 2 hops of each other can use same color. Further, every node can have at the most two children. In each time period $T$, a node can forward only one packet (this could be its own packets or on behalf of one of its two children). The parent can signal which child should send a packet next by ordering the child IDs in the Harvest message, as we shall explain later.

To begin the TDMA scheduling, the base station selects a color for itself and starts sending beacon messages in its timeslot. The 1-hop neighbors of base station randomly select an available color and start sending their payload. Every node's packet contains the IDs of its 1-hop neighborhood transmitters. If node A hears its 1-hop neighbor transmitting in the same time-slot then one of two nodes gives up its color. The priority among the contending nodes is decided by considering which nodes was the first to select the color and then by the IDs of the nodes. Thus, priority is locally computed by looking at the sequence number of the messages and the unique IDs of the nodes.

The underlying MAC layer in Harvest is CSMA/CA based. As a result, even if two nodes try to transmit in the same time slot, only one of them can succeed. We claim that this phenomenon applies to both scenarios, viz. when the two contending nodes are 1-hop neighbors of each other or 2-hops neighbors of each other. The nodes in the 1-hop neighborhood of the base station select a color and the wave propagates. After a node selects a color for itself or finds that there are no available colors in its D-2 neighborhood, it turns off the backoffs in the underlying CSMA/CA. Every node maintains a list of node IDs, which are using the 4 colors in its D-2 neighborhood, as a soft state. Whenever a node finishes its data transmission, it stops transmitting and its color is available for reuse. All the nodes in the D-2 neighborhood learn this information by the virtue of the soft state and enable backoff in the underlying CSMA/CA. And the process of randomized color selection repeats.

The selection of 2 senders and the color assignment is a distributed operation. The operation is initiated from a single base station as opposed to the nodes in the network, since uniquely selecting nodes in a distributed manner would incur additional message overhead for coordination. The approach outperforms a centralized solution because in the latter a single node (such as the base station) would need to collect the entire topology information to compute disjoint paths between two nodes. Further, the operation would have to repeated whenever a new sender is selected.

\vspace{0.15cm}
\noindent 
\textbf{Data Forwarding Protocol.}~~~As soon as a node has selected its parent and uniquely selected a color in its D-2 neighborhood, it starts sending data packets to its parent in the corresponding timeslot. A parent can choose to receive packets from either of its children. If a parent node has 1 buffer space, then it can receive only 1 packet in the entire time period $T = 4*t_{S}$ ($t_{S}$ is long enough to transmit a Harvest message with CSMA/CA backoffs disabled). In this csae, a parent node receives packets from a child in every alternate time period. The process of alternation ensures that the colors assigned to its children are not unassigned. This is a minor variation in the Harvest protocol as described above. Instead of one packet buffer at each node, more than one packet buffers can be allocated at each node. This will expedite the data collection.

\vspace{0.15cm}
\subsection{Implementation Description}

\begin{figure}[htbp]
	\centering
		\includegraphics[width=3in]{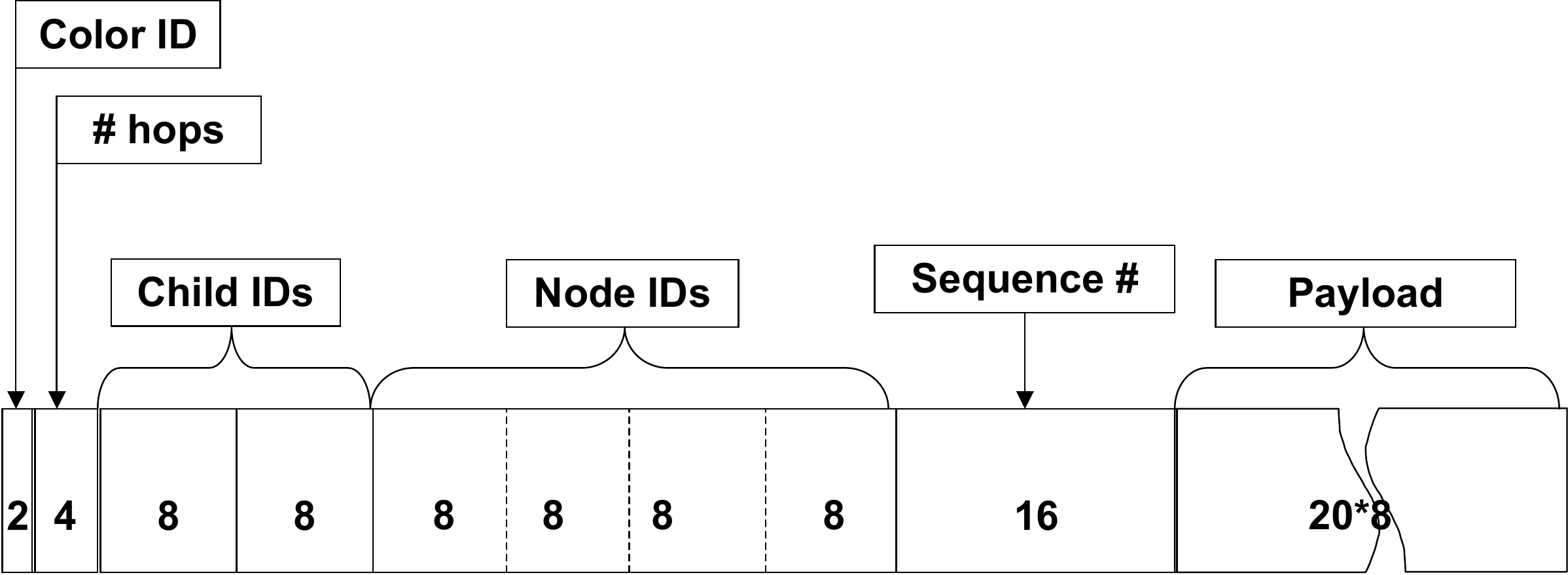}
		\caption{Harvest Message Structure}
		\label{'fig:message_stucture'}
\end{figure}

In this section, we describe the implementation of Harvest in NesC under TinyOS 1.x release. Harvest has a single message structure; Figure \ref{'fig:message_stucture'} illustrates this structure, using numbers that denote field sizes in terms of bits. 

The payload in each Harvest packet is 20 bytes (this is in contrast to Straw's payload of 22 bytes). The color ID identifies one of the 4 colors used by the node. The \# hops denotes the distance of the sender from the base station. This information is used by a node to select it parent, which has the minimum distance to the base station, among the set of nodes within 1-hop neighborhood.

The child IDs are used identify the IDs of the sender's children. A non-null value declares that the sender is available for forwarding. Also, the sender can use this field to declare its decision about the selected children in case multiple node are in contention for the selection. In particular, the ID in the first field among the two, should send packet in the next time period.

The array of 4 node IDs denotes the IDs of the 4 nodes, in the sender's D-2 neighborhood, which are currently using the 4 colors. The array is ordered in the increasing order of the color IDs. Every node copies the array received from its 1-hop neighborhood and maintains it as a soft state. If a color is not refreshed for a certain time, then the node assumes that the color is free and sends this information as part of its messages. The sequence number is a monotonically increasing number and denotes the sequence number of the packet. It is used in the calculation of WMEWMA link estimate and to select a unique node in the case of 2 nodes contending for the same color or same parent. The range of sequence numbers can be chosen depending upon the number of packets to be transmitted and also the number can be recycled to save space.

There are no explicit sender ID and the destination ID fields. The sender ID can be retrieved from the message by reading the node ID at the location of sender's color ID. The destination ID field is used from the TOS header in the TinyOS packet. Harvest uses promiscuous mode of transmission so that neighboring nodes can learn about the color allocation. But only the node identified as the destination node forwards the packet to the base station. The base station's ID is 0. The receiver can identify whether the message is from the base station or not by looking at the sender ID. 

Harvest does not need an explicit tine synchronization service for its TDMA to function. Every packet contains the D-2 color of the sender. We use the synchronous reception property of the wireless medium to achieve time synchronization among the nodes \cite{je:ts}. In particular, when a node hears a packet from its parent, it synchronizes its time with that of its parent. Since base station is the root of the tree, all the time at all the nodes is synchronized to that of the single clock of the base station by virtue of induction. This synchronization scheme is also used in Sprinkler \cite{vn:sprinkler_rtss}, which uses TDMA.

\vspace{0.15cm}
\section{Performance Evaluation}
In this section, we evaluate the latency and number of packet transmissions for Harvest's randomized slot assignment algorithm and data collection protocol.

\vspace{0.15cm}
\subsection{Randomized Slot Assignment Performance}

As described earlier, Harvest uses an underlying CSMA/CA protocol for color selection.  In particular, when a node receives a message, it finds out the available colors in its D-2 neighborhood from the received message. If one or more colors are available, the node randomly selects an available color and starts transmitting from the corresponding TDMA slot in next time interval. It is possible that two or more nodes can simultaneously select the same color and therefore their transmitted packets can collide with each other. We show here that in O$(1)$ time, a unique node will select a unique color in the node's D-2 neighborhood.

TinyOS uses a variant of non-persistent CSMS/CA protocol \cite{aw:tinyos_mac}. We briefly recall the definition of non-persistent CSMA/CA protocol \cite{db:data_networks}:
\begin{enumerate}
	\item A node senses channel before transmission.
	\item If the channel is free, it immediately transmits a frame; otherwise it waits for a random amount of time.
	\item After waiting, it repeats step 1.
\end{enumerate}
In the case of TinyOS, a node waits for a random amount of time before it senses the channel. This ensures that the transmissions are not synchronized. Because of the initial random wait, the throughput of the CSMA/CA in TinyOS is better than that of the classical non-persistent CSMA/CA.
\vspace{0.15cm}

\begin{theorem}
Given that the degree ($\Delta$) of network is bounded, Harvest takes O$(1)$ time for assigning a unique color to a node in its D-2 neighborhood.
\end{theorem}
\vspace{0.15cm}

\begin{proof}
Given that $\Delta$ is bounded, for non-persistent CSMA/CA, there exists a constant $\tau > 0$ such that the probability of a frame transmission without collision is at least $\tau$ \cite{db:data_networks}. The same holds for the CSMA/CA in TinyOS which is variant of non-persistent CSMA/CA. 

Therefore, the expected time for a frame transmission without failure is also O$(1)$. In the event that two or more nodes select the same color and transmit in the same timeslot, in O$(1)$ time, a unique node will succeed in a transmission without failure. After one transmission without collision, all the nodes in the D-1 neighborhood will learn that the color is not available. Similarly, for the D-2 neighborhood, a unique color is selected in O$(1)$ time since the packet delivery rate between D-2 neighbors is non-zero. After one transmission without collision by the successful node, or via a neighbor of the successful node, the color assignment of the successful node get conveyed to its 2-hop neighborhood.
\typeout {CHECK LAST STATEMENT}
\end{proof} 

\begin{figure}[htbp]
	\centering
		\includegraphics[width=3in]{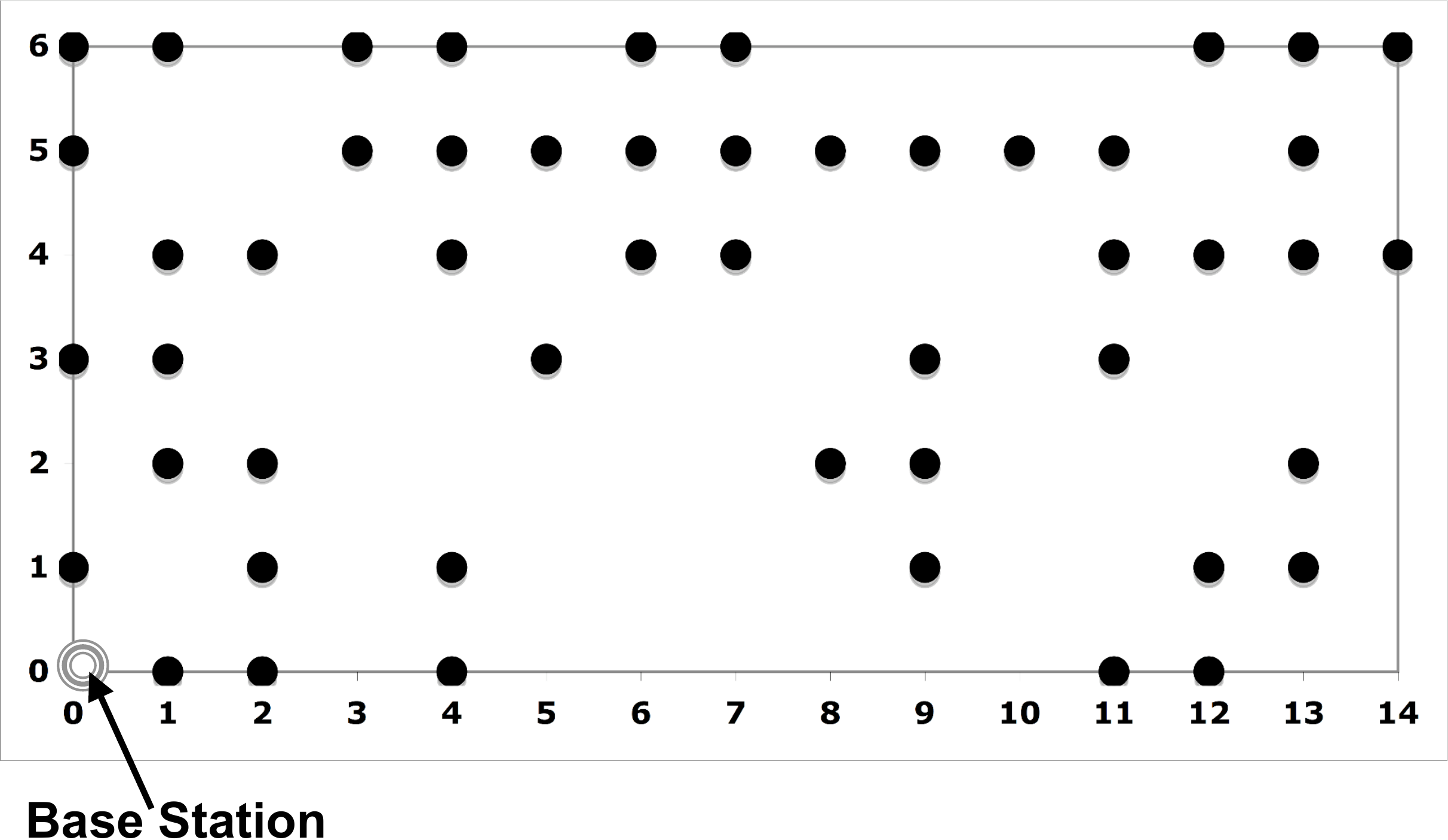}
		\caption{Testbed Topolgy }
		\label{'fig:testbed_topology'}
\end{figure}

The value of $\tau$ depends upon the range of values for random wait and $\Delta$. Instead of finding the value of $\tau$, we perform experiments to measure the convergence time of Harvest's slot assignment for different values of $\Delta$. We use 51 XSM motes in an indoor testbed, Kansei. An XSM mote uses Chipcon's CC1000 radio and is for the purposes of this experiment similar to a Mica-2 mote. We use the TinyOS 1.x release and the standard MAC that comes with the 1.x release. The topology of the network is as shown in the Figure \ref{'fig:testbed_topology'}. The motes are placed in grid with 3ft unit separation on the X and Y axes. We uses default power level and default frequency for transmission. The mote at location (0,0) is selected as the base station, as shown in Figure \ref{'fig:testbed_topology'}. Each slot is of the duration 31 msec, which is the minimum possible given that the radio transmission takes at 23 msec and the UART transmission takes at least 8 msec in TinyOS 1.x over XSM. Each node has a payload of 100 packets to be sent to the base station.

We measure the time required to collect all the data packets from the first mote after the start of the experiment as a function of the number of nodes. The time is sampled at a granularity of 30 times the time for a transmission period. The number of sampling periods required for the first node to complete data exfiltration denotes the convergence time of the color selection algorithm. As shown in Table \ref{"table:scalbility_of_color_selection"}, the convergence time has a variance of 1 sampling period, which is negligible. Hence, for the non-persistent CSMA/CA implementation in TinyOS 1.x release, the convergence time of  Harvest's randomized slot assignment is negligible for a $\Delta$ up to 51.

\begin{table}[htp]
\centering
\begin{tabular}{|c|c|} \hline 
\textbf{\# nodes}	&	\textbf{Convergence time}\\ \hline
6	&	8\\ 
12	&	8\\
18	&	8\\
22	&	8\\
31	&	7\\ 
42	&	8\\
51	&	9\\ \hline
\end{tabular}
\caption{Scalability of color selection}
\label{"table:scalbility_of_color_selection"}
\end{table}

\vspace{0.15cm}
\subsection{Data Collection Protocol}

\vspace{0.15cm}
\noindent \textbf{Latency.}~~~We define the total latency of Harvest data collection to be the duration between the moment that the base station receives the first data packet and the moment it receives the last data packet. Since the base station has 2 children and there are 4 timeslots per time period $T, T = 4*t_{s}$, it receives 2 packets per time $T$. Therefore, for $n$ nodes and $M$ number of packets from each node, the time required to receive $n*M$ packets is $n*M*2*t_{s}$, which is O$(n*M)$. The time required to build the tree rooted at the base station is in the worst case O$(n)$. Note that the tree building is happening in parallel to the data collection. But for the worst case analysis, we can assume that the two processes happen sequentially. In that case, the total latency of Harvest is O$(n)$ + O$(n*M)$ = O$(n*M)$.

\vspace{0.15cm}
\noindent \textbf{Number of Transmissions.}~~~Let $h$ be the average height of a node in the routing tree. Therefore given $n*M$ packets, the total number of transmissions is O$((n*M*h)/2)$.

\vspace{0.15cm}
\section{Performance Comparison}
\label{'section:performace_results'}

\vspace{0.15cm}
\subsection{The Straw Protocol}
In this section, we compare the performance of Harvest with that of Straw \cite{sk:straw}. Similar to Harvest, the objective of Straw protocol is to collect bulk data from all the nodes at the base station. Unlike Harvest, Straw collects data from one node at a time. For each node, the data collection is divided into two phases, viz. broadcast and collection. In case the collection phase loses packets, the two phases are repeated to recover from loss. (The broadcast command in the recovery phase contains the sequence numbers of the lost packets.) The overall goal of the protocol is to minimize latency and number of packet transmissions. The broadcast phase disseminates the ID of a selected node, from which data is to be collected. Following the broadcast phase, the selected node periodically sends packets to the base station. The route is selected using MintRoute protocol.

For all nodes that are at a distance greater than 2-hops from the base station, the transmission period in Straw is $3*t_{h}$, where $t_{h}$ is the time required to traverse single hop. The number 3 is chosen to reduce the interference with data forwarding at an upstream node. If we color the nodes that transmit at the same time, then the coloring of transmitting nodes effectively yields a D-2 coloring. Note that the transmitting nodes induce a one dimensional graph, in other words, a single line (and hence the name ``straw''). Due to the fact that each node sends packets at the period of $3*t_{h}$, the base station receives a packet after every $2*t_{h}$ time. For nodes at 1-hop and 2-hop distances from the base station, the transmission period is $t_{h}$ and $2*t_{h}$ respectively.

The initial broadcast command sets up the colors on the linear path from a node to the base station. This corresponds to a deterministic slot assignment, as compared to the randomized slot assignment of Harvest. Further, a node from which data is collected, is selected by the base station as opposed to the local, distributed selection in Harvest. 

\vspace{0.15cm}
\subsection{Latency Comparison}
\vspace{0.15cm}
\subsubsection{Theoretical Comparison}
\label{'subsubsection:theoretical_comparison_latency'}
Straw uses a broadcast for slot assignment. In each broadcast phase, a node forwards the broadcast command once. For collecting data from $n$ nodes, Straw will therefore employ $n$ broadcast sessions, on average lasting for at least O($h$) time. Therefore, the total latency for assigning slots is O$(n*h)$ as compared to O$(n)$ for Harvest. 

In Straw's data collection protocol, only the nodes on the path from the current sender to the base station are transmitting. The rest of the network is idle, in other words, spatial reuse is limited. If the rest of the network lies outside interference distance from the the transmitters, then an idle node from the rest of the network can send its data towards the base station. However, finding a nodes outside interference distance from the current transmitters could be impossible, especially near the base station. A solution is to increase the number of D-2 colors from 3. 

Instead of a linear structure, Harvest utilizes a tree structure to collect data packets. Given the concurrency constant of 2, Harvest uses a binary tree. Harvest uses 4 colors in order to ensure that the binary tree can be D-2 colored. In that case, the base station receives 2 packets every $4*t_{s}$, where $t_{s}$ is the duration of a timeslot. Therefore the rate of data collection at the base station is equal to a packet after every $t_{s}$ time. Note that we can utilize any m-ary tree and $C$ colors, and the resulting rate of data collection at the base station would be $m/C$.

In Straw, the rate of data collection from the nodes at more than 2-hops from the base station is 1 packet per $3*t_{s}$. If we assume that the number of nodes at 1-hop and 2-hop distance from the base station is far less than that the total number of nodes $n$, the latency of data collection for Straw is $n*M*3*t_{s}$. Therefore, data collection of Harvest has 33.33\% lower latency than that of Straw.

Further, the overall order complexity of the latency of Straw is O$(n*h)$ + O$(n*M)$, which exceeds the O$(n*M)$ of harvest if O$(h)$ is greater than O$(M)$.
\vspace{0.15cm}

\vspace{0.15cm}
\subsubsection{Simulation-based Comparison}
\label{'subsubsection:simulation-based_comparison_latency'}
\begin{figure}[htbp]
	\centering
		\includegraphics[width=3in]{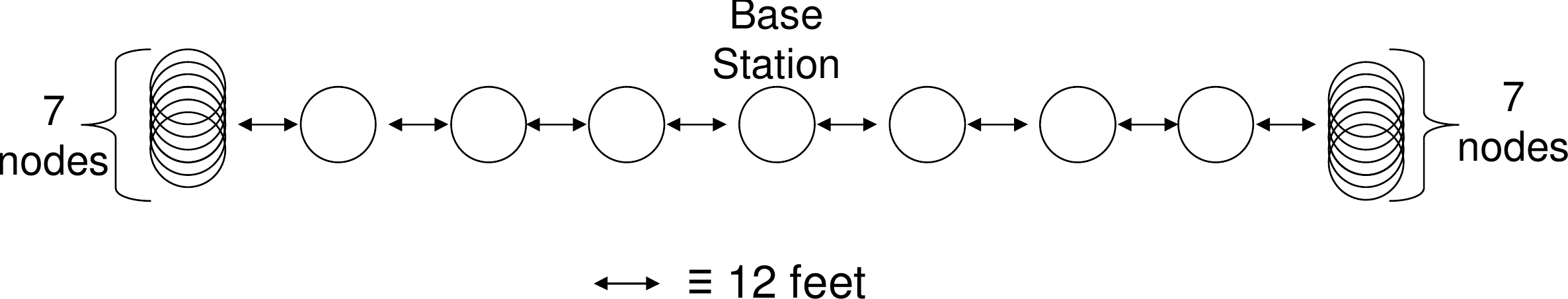}
		\caption{Network topology for simulation}
		\label{'fig:21-nodes_spacing-8_topology'}
\end{figure}

To validate the claimed improvement in latency, we perform simulations in TOSSIM. We setup a network of 20 non-base station motes and 1 base station node. As shown in Figure \ref{'fig:21-nodes_spacing-8_topology'}, we ensure that there are nodes at more than 2-hop distance from the base station. Also, the base station has more than 1 node at 1-hop distance.

The QueuedSend buffer module at the TinyOS's MAC layer uses explicit acknowledgment. In the case of unsuccessful transmission, a retransmission is attempted. However, the retransmission could happen in an incorrect timeslot, resulting in a collision. Therefore, we have disabled the MAC layer ACK in this simulation. However, we can still utilize the MAC-level ACK by channeling the ACK information to the Straw and Harvest protocol layer. Use of ACK will increase the reliability to data collection. 

All of the simulation are done under TOSSIM. This NesC-code simulator has an option to instantiate a link quality set given the node placement. The links qualities are based on some empirical measurements carried out for MICA-2 motes. The links qualities vary in spatial and temporal dimensions. Since the current implementation of Harvest assumes symmetric links and base its parent selection criteria by measuring link quality in one direction, we have pre-processed the link quality set so that all the links are symmetric. In our future work, we will refine the implementation to deal with asymmetric links by computing link quality in both directions. In particular, link quality from the child to parent will be computed by counting the number of successful and failed ACKs. We use our NesC implementation and we use the Straw code which has been available as part of a Golden Gate Bridge health monitoring project contribution folder under TinyOS 1.x release. 

We measure the rate at which data is collected at the base station. We observe that the rate of data collection is $1.67$ packets per $4*t_{s}$ for Harvest and $0.8$ packets per $3*t_{s}$ for Straw, as shown in Table \ref{"table:latency_comparison"}. The rate of data collection is lower than the respective theoretical values due to the fact that ACKs are disabled. The observed latency gain under simulation is $36\%$, which is close to the theoretical value of $33.33\%$.

\begin{table}[htp]
\centering
\begin{tabular}{|c|c|c|} \hline 
\textbf{Service}	&	\textbf{Theoretical}	&	\textbf{Simulation}\\ \hline
Straw	&	1 packets/$(3*t_{s})$	&	0.8 packets/$(3*t_{s})$\\ 
Harvest	&	2 packets/$(4*t_{s})$	&	1.67 packets/$(3*t_{s})$\\ \hline
\end{tabular}
\vspace{0.15cm}
\caption{Latency Comparison}
\label{"table:latency_comparison"}
\end{table}

\vspace{0.15cm}
\subsection{Energy Comparison}

\vspace{0.15cm}
\subsubsection{Theoretical Comparison}
Straw uses a broadcast to disseminate the command to send the ID of a selected node. This is equivalent to the slot assignment in Harvest. In a broadcast phase, each node in the network forwards a newly heard packet exactly once. Therefore, the total number of transmissions in a broadcast phase are $n$, where $n$ is the total number of nodes in the network. Therefore, to collect data from $n$ nodes, the total number of packet transmissions are $n^{2}$. 

Also, each broadcast phase is followed by a reply from a selected node to the base station. The total number of transmissions, for each reply, is a function of number of hops from the selected node to the base station. In the worst case, the average path length in the network could be $n/2$. In that case, the total number of replies for all nodes is $n^{2} + n/2$, which is O$(n^{2})$. 

In Harvest, the control information pertaining to slot assignment is piggybacked on the data messages. Therefore, Harvest does not have packet transmissions for slot assignment. Hence, it saves O$(n^{2})$ number of packet transmissions as compared to Straw. 

We assume that Straw and Harvest both use the shortest path routes to transmit data packets to the base station. In that case, the total number of data packet transmissions for data collection purposes are the same for Straw and Harvest.  In particular, this number is O$((n*M*h)/2)$.

The total number of messages for Straw is O$(n^{2})$ + O$((n*M*h/2)$. 


\vspace{0.15cm}
\subsubsection{Simulation-based Comparison}
In reality, the radio behavior is more complex than that represented by the simplistic unit-disk radio model. Not only is packet delivery rate less than 100\% but it also varies in space and time. Therefore, we conduct simulations over a multi-hop network to compare the number of packet transmissions for Straw and Harvest. We conduct simulations in the same network topology as used in Section \ref{'subsubsection:simulation-based_comparison_latency'}. In future, we plan to compare results in a real sensor network.

\begin{figure}[htbp]
	\centering
		\includegraphics[width=3in,height=1in]{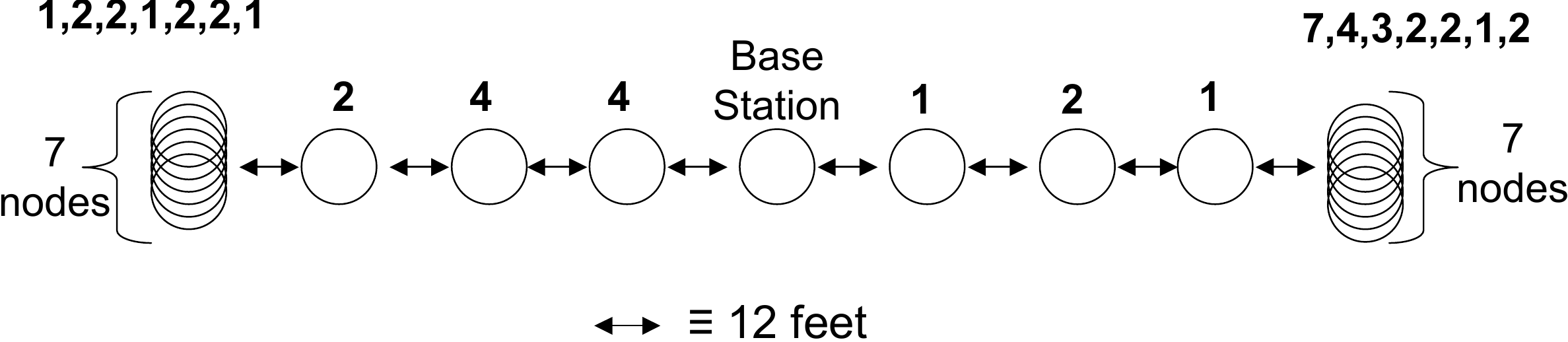}
		\caption{Number of broadcast sessions for 21 nodes in Straw}
		\label{'fig:energy_comparsions'}
\end{figure}

The number on top of each node, in Figure \ref{'fig:energy_comparsions'}, illustrates the number of broadcast sessions required to reliably convey the command to each of the 20 nodes. The total number of broadcast sessions are 46. Given 20 nodes, Straw consumes 20 times 46, i.e. 920, more packet transmissions. 

\vspace{0.15cm}
\section{Harvest Extensions and Discussion}

\vspace{0.15cm}
\noindent 
\textbf{Duty Cycling of Radios.}~~~As we discussed in Section 2, an idle radio draws a significant amount of current and so energy efficiency is gained by letting idle nodes sleep. In Harvest, we achieve this as follows. When a node sees that no colors are available for itself in its interference region (i.e., its 2-hop neighborhood), it can switch off its radio until a color is expected to be available again. Given some knowledge of the number of packets to be transmitted that color and by observing the sequence number of the packet currently being transmitted for that color, a sleeping duration can be readily calculated.

Furthermore, once a node is done with its role in the convergecast, it can switch off its radio permanently. A node is defined to be done with its transmissions after it has sent all of its packets and the packets of its children.

\vspace{0.15cm}
\noindent 
\textbf{Reliability when all Children Transmit Concurrently.}~~~As described in Section 3, the data collection protocol allows non-base station nodes to forward data from one or more of its children. (If more than one child can transmit, then the protocols maintains at least one buffer per child.) When only one child is allowed to transmit, the implicit acknowledgement scheme suffices for nodes to discover whether or not their transmissions were successfully received. 

When more than one child is allowed to transmit, using the implicit acknowledgement scheme implies either a delay in loss detection or a modification of the protocol to expose more node buffer information. One alternative in this case would be to use explicit acknowledgements. If we assume that explicit acknowledgements can be send immediately (or within some constant delay after message reception), then the tranmission slots can be extended to subsume both the transmission time and the acknowledgement time.

\vspace{0.15cm}
\noindent 
{\bf Continuous Streaming of Data to the Base Station}.~~~Harvest collects data simultaneously from multiple nodes, as opposed to receiving data from only one node at a time. In this sense, the data received at the base station resembles a continuous stream of data from the network ordered in time. It is therefore conceivable to use Harvest as the basis for collecting in an on-line fashion continuous data streams from the network. 

Note that the description in Section 3 for the case where all nodes forward data from multiple children allows the possibility that the data from each child is forwarded in a round robin. More sophisticated rules for fair scheduling that consider the distance of the node from the base station can be defined to achieve global fairness, otherwise nodes near base station will contribute more packets as opposed to the ones farther from the base station. One extension of Harvest that we are presently studying is in the context of real-time wireless sensor network applications, such as visualizing link quality of the network in real time or viewing consistent global snapshots of the wireless sensor network.

\section{Related Work}

\vspace{0.15cm}
\noindent
{\bf TDMA and CSMA}.~~~
Herman et al \cite{th:distributed_tdma} have proposed a randomized TDMA algorithm that first forms clusters, each with a unique cluster-head. Each cluster-head then allocates colors to its children. Cluster-heads are ordered in a monotonically increasing order, so the color assignment occurs sequentially per that order. A similarity between this work and Harvest's TDMA scheduling is the use of underlying CSMA/CA MAC layer to CSMA/CA to communicate control information pertaining to node coloring and TDMA scheduling. Z-MAC \cite{ir:zmac} uses both TDMA and CSMA/CA features in manner different from Harvest. Z-MAC is hybrid MAC that uses TDMA under high contention and CSMA under low contention, whereas data transmission in Harvest is always in TDMA mode. Kulkarni and Arumugam \cite{sk:sstdma} describe TDMA based protocols that are optimized for convergecast, that work however assumes grid localization.

RID \cite {gz:rid} is a radio interference detection service that detects interference relations between nodes at run-time, and provides higher fidelity for collision avoidance when using TDMA. The RID approach would be a suitable candidate for enabling Harvest's distributed coloring protocol.

\vspace{0.15cm}
\noindent
{\bf Convergecast routing.}~~There is a rich body of work on convergecast routing for wireless sensor systems. Several protocols assume location information. Most of the others such as MintRoute \cite{aw:mintroute}, RMST \cite{ir:rmst}, PSFQ \cite{cw:psfq}, Drain \cite{gt:snms_design} are, unlike Straw and Harvest, not optimized for the energy and latency requirements associated with the collection of payloads that can well be in the thousands of packets per node. For instance, MintRoute does not pipeline transmissions, which would yield higher latency for bulk data transport, and Drain is optimized for the case of a single packet payload per source mote. 

\vspace{0.15cm}
\noindent
{\bf Reliability.}~~~The study of reliability in convergecast has often arisen in the context of concurrent event detections, which tend to occur in a bursty manner or with multiple sources are continuously/periodically generating packets (with low duty cycle). RBC \cite{hz:rbc} focuses on the former whereas the traffic models considered in CODA \cite{cw:coda} and ESRT \cite{ys:esrt} focus on the latter.  RBC deals with bursts by maintaining information about queue conditions of the neighboring node as well as number of times enqueued packets were retransmitted, which results in sizeable RAM usage. Also, the queue condition has to be transmitted in RBC, which results in sizeable communication overhead.   
The alternative approaches of packet retransmissions, of acknowledgements, of hop-by-hop recovery, as well as selecting alternative routes upon link failure are also relevant approaches for improving the reliability of Harvest in particular application contexts.   The use of TDMA and receiver-driven flow control mitigate the consideration of congestion. 

Coding of bulk data is a relevant approach for tolerating packet loss in bulk convergecast. Kim et al \cite{sk:secon} have considered the use of erasure codes.  We have regarded this relevant consideration as being orthogonal to the pipelining and spatial reuse considerations of Harvest.

\vspace{0.15cm}
\section{Conclusion}
We have presented a bulk data collection service, Harvest, for energy constrained wireless sensor nodes. Harvest assumes a bounded node density, i.e., degree $\Delta$. This assumption enable us to
We assign distance-k (our exposition has used k=2) colors to nodes in O$(1)$ time by utilizing an underlying CSMA/CA MAC layer. We use a constant number of colors in the entire network, which enables the per node computation of its TDMA schedule to occur in O$(1)$ time. Harvest exploits the spatial parallelism in collecting data, thereby achieving a latency gain of at least 33\% in large networks (i.e., networks with more than three hops) as compared to that of Straw. Harvest also avoids the O$(n^2)$ number of broadcasted control transmissions used in Straw. Further, Harvest requires only O$(1)$ number of buffers at each node. Therefore, Harvest is suitable for large scale network of wireless sensor network. We provide theoretical bounds on the performance of Harvest and perform simulation results to validate the theoretical bound. We find that the spatial parallelism not only reduces latency, but also creates an opportunity to collect global data in a fair real-time manner. Our present work is studying extensions of Harvest for the case of on-line continous data streaming from the network to the base station.

\bibliographystyle{plain}
\bibliography{my_bibliography}

\begin{thebibliography}{10}

\bibitem{db:data_networks}
D.~Bertsekas and R.~Gallager.
\newblock {\em Data Networks}.
\newblock Prentice Hall, Englewood Cliffs, NJ, 1987.

\bibitem{je:ts}
J.~Elson.
\newblock {\em Time Synchronization in Wireless Sensor Network}.
\newblock PhD thesis, UCLA, 2003.

\bibitem{th:distributed_tdma}
T.~Herman and S.~Tixeuil.
\newblock A distributed tdma slot assignment algorithm for wireless sensor
  networks.
\newblock In {\em Algorithmic Aspects of Wireless Sensor Networks}, pages
  45--58, 2004.

\bibitem{sk:straw}
S.~Kim.
\newblock Wireless sensor networks for structural health monitoring.
\newblock Master's thesis, University of California at Berkeley, USA, 2005.

\bibitem{sk:secon}
S.~Kim, R.~Fonseca, and D.~Culler.
\newblock Reliable transfer on wireless sensor networks.
\newblock In {\em Annual IEEE Communications Society Conference on Sensor and
  Ad Hoc Communications and Networks}, 2004.

\bibitem{sk:channel_models}
S.~Krumke, M.~Marathe, and S.~Ravi.
\newblock Models and approximation algorithms for channel assignment in radio
  networks.
\newblock {\em Wireless Networks}, 7(6):575--584, 2001.

\bibitem{sk:sstdma}
S.~Kulkarni and M.~Arumugam.
\newblock {\em SS-TDMA: A Self-Stabilizing MAC for Sensor Networks}, chapter In
  Sensor Network Operations.
\newblock IEEE Press, 2005.

\bibitem{vn:sprinkler_rtss}
V.~Naik, A.~Arora, P.~Sinha, and H.~Zhang.
\newblock Sprinkler: A reliable and energy efficient data dissemination service
  for wireless embedded devices.
\newblock In {\em The 26th IEEE Real-Time Systems Symposium}, December 2005.

\bibitem{ir:zmac}
I.~Rhee, A.~Warrier, M.~Aia, and J.~Min.
\newblock Z-mac: A hybrid mac for wireless sensor networks.
\newblock In {\em SenSys}, pages 90--101, 2005.

\bibitem{ys:esrt}
Y.~Sankarasubramaniam, O.~Akan, and I.~Akyildiz.
\newblock Esrt: Event-to-sink reliable transport in wireless sensor networks.
\newblock In {\em The ACM Symposium on Mobile Ad Hoc Networking and Computing
  (MobiHoc)}, 2003.

\bibitem{ir:rmst}
F.~Stann and J.~Heidemann.
\newblock Rmst: Reliable data transport in sensor networks.
\newblock In {\em The 1st IEEE Intl. Workshop on Sensor Network Protocols and
  Applications (SNPA)}, pages 102--112, 2003.

\bibitem{gt:snms_design}
G.~Tolle and D.~Culler.
\newblock Design of an application-cooperative management system for wireless
  sensor networks.
\newblock In {\em Second European Workshop on Wireless Sensor Networks}, 2005.

\bibitem{cw:psfq}
C.~Wan, A.~Campbell, and L.~Krishnamurthy.
\newblock Psfq: A reliable transport protocol for wireless sensor networks.
\newblock In {\em WSNA '02: Proceedings of the 1st ACM International Workshop
  on Wireless Sensor Networks and Applications}, pages 1--11, New York, NY,
  USA, 2002. ACM Press.

\bibitem{cw:coda}
C.~Wan, S.~Eisenman, and A.~Campbell.
\newblock Coda: congestion detection and avoidance in sensor networks.
\newblock In {\em SenSys}, pages 266--279, 2003.

\bibitem{aw:tinyos_mac}
A.~Woo and D.~Culler.
\newblock A transmission control scheme for media access in sensor networks.
\newblock In {\em ACM/IEEE International Conference on Mobile Computing and
  Networking (MobiCom)}, pages 221--235, 2001.

\bibitem{aw:mintroute}
A.~Woo, T.~Tong, and D.~Culler.
\newblock Taming the underlying challenges of reliable multihop routing in
  sensor networks.
\newblock In {\em SenSys '03: Proceedings of the 1st international conference
  on Embedded networked sensor systems}, pages 14--27, 2003.

\bibitem{hz:rbc}
H.~Zhang, A.~Arora, Y.~Choi, and M.~Gouda.
\newblock Reliable bursty convergecast in wireless sensor networks.
\newblock In {\em 6th ACM International Symposium on Mobile Ad Hoc Networking
  and Computing}, 2005.

\bibitem{gz:rid}
G.~Zhou, T.~He, J.~Stankovic, and T.~Abdelzaher.
\newblock Rid: radio interference detection in wireless sensor networks.
\newblock In {\em The 24th Annual Joint Conference of the IEEE Computer and
  Communications Societies (INFOCOM)}, pages 891-- 901, 2005.

\end{thebibliography}

\end{spacing}

\end{document}